%Paper: hep-ph/9309226
%From: JFGUCD@UCDHEP.UCDAVIS.EDU
%Date: Fri, 3 Sep 1993 17:51 PDT

\def\delrho{\Delta \rho}

\def\emem{e^-e^-}

\def\wpm{W^{\pm}}

\def\mt{m_t}

\def\hawaii{{\it Proceedings of the 2nd International Workshop on
``Physics and Experiments with Linear $\epem$ Colliders''}, ed. F. Harris,
Waikoloa, HI, April 26-30, 1993}

\def\gam{\gamma}

\def\gev{~{\rm GeV}}
\def\tev{~{\rm TeV}}

%%Journal definitions

\def\prdj#1{{\it Phys. Rev.} {\bf D{#1}}}

\def\plbj#1{{\it Phys. Lett.} {\bf B{#1}}}

\def\mt{m_t}
\def\wp{W^+}
\def\wm{W^-}
\def\rta{\rightarrow}

\def\mw{m_W}
\def\mz{m_Z}
\def\anti{\overline}

\def\ifmath#1{\relax\ifmmode #1\else $#1$\fi}
\def\half{\ifmath{{\textstyle{1 \over 2}}}}

\def\quarter{\ifmath{{\textstyle{1 \over 4}}}}

\def\3quarter{{\textstyle{3 \over 4}}}
\def\third{\ifmath{{\textstyle{1 \over 3}}}}

\def\fivethirds{{\textstyle{5 \over 3}}}
\def\eightthirds{{\textstyle{8 \over 3}}}

\input phyzzx
\Pubnum={$\caps UCD-93-29$}
%\pubtype={}
\date{August, 1993}
%\Pubnum={}
%\date{}

%\nopagenumbers
\titlepage
\vskip 0.75in
\baselineskip 14pt
\hsize=6in
\vsize=8.5in
\centerline{{\bf HIGGS SECTOR MOTIVATIONS FOR AN $\bf e^-e^-$ LINEAR COLLIDER}
\foot{To appear in \hawaii.}
}
\vskip .5in
\centerline{ J.F. Gunion}
\vskip .075in
\centerline{\it Davis Institute for High Energy Physics,
Dept. of Physics, U.C. Davis, Davis, CA 95616}

\vskip .075in
\centerline{\bf Abstract}
\vskip .075in
\centerline{\Tenpoint\baselineskip=12pt%\parindent=1pc
\vbox{\hsize=12.4cm
\noindent I briefly review the crucial role an $\emem$ linear collider
could play in unravelling the nature of a non-minimal Higgs sector
and/or strongly-interacting $WW$ sector..
}}

\vskip .15in
\noindent{\bf 1. Introduction}
\vskip .075in

Much has been written
\Ref\gunlcws{See, for instance, J.F. Gunion, {\it Higgs Bosons at
an $\epem$ Linear Collider: Theory and Phenomenology}, these proceedings.}
about the virtues of a linear $\epem$ collider for probing the Higgs
sector or, more generally, revealing  the nature and source of electroweak
symmetry breaking (EWSB).  While an $\epem$ collider is fully adequate
as a probe of Higgs physics in the case of the minimal Standard Model
(MSM), it generally has important limitations if the Higgs sector is
non-minimal or if the $W$ ($W\equiv\wpm,Z$) boson sector is strongly
interacting.  For a perturbative non-minimal Higgs sector, $\emem$
collisions could be crucial for observing doubly-charged Higgs
bosons via $\emem\rta \nu\nu\wm\wm\rta \nu\nu H^{--}$ as well as
various exotic couplings of neutral and singly-charged Higgs bosons.
If the $W$ boson sector is strongly interacting, only the combination
of $\epem$ and $\emem$ collisions will allow a full
investigation of all $WW$ scattering channels, as required
to fully understand electroweak symmetry breaking.
Here I review the basic theoretical ideas and phenomenology that
provide significant motivation for retaining an $\emem$ collision
option at the next linear collider (NLC).

\smallskip
\noindent{\bf 2. Perturbative Higgs Sector Extensions}
\smallskip

In the context of perturbative theories
containing elementary Higgs bosons, the MSM need not be nature's
choice. Many generalizations have been  discussed.
\Ref\hhg{For a review see
J.F. Gunion, H.E. Haber, G.L. Kane, and S. Dawson, {\it The Higgs Hunter's
Guide}, Addison-Wesley, Redwood City, CA (1990).}\
Here, I focus entirely on extensions of the Higgs sector only.
Supersymmetric generalizations of the models to be discussed
are certainly possible, but will not be considered here.

There are two crucial conditions that must be satisfied by any
perturbative Higgs sector generalization:
\point no problems with unitarity at high energy;
\point no violation of $\rho\simeq 1$.

Turning first to $\rho$, we recall the tree-level result:
$$\rho={\sum_{T,Y}\left[4T(T+1)-Y^2\right]\left|V_{T,Y}\right|^2 c_{T,Y}
\over \sum_{T,Y} 2Y^2 \left|V_{T,Y}\right|^2}\,\eqn\rhoform$$
where $\langle\phi(T,Y)\rangle=V_{T,Y}$ defines the vacuum expectation value
of the neutral Higgs field member of the Higgs representation with total
$SU(2)_L$ isospin and hypercharge specified by $T$ and $Y$, respectively,
and $c_{T,Y}$ is 1 for a complex representation and $1/2$ for
a $Y=0$ real representation. For a single representation to yield
$\rho=1$ we require $(2T+1)^2-3Y^2=1$, which is satisfied for any number of
Higgs doublets ($T=1/2$, $Y=\pm 1$), to which any number
Higgs singlets ($T=Y=0$) can be added.

The MSM employs a single Higgs doublet.
The most attractive extension is to two Higgs doublet fields.
This extension, including its supersymmetric generalization is
reviewed at length in Ref.~\gunlcws. In general, $\emem$ colliders
are not particularly critical to fully explore such a generalization.
However, higher representations should certainly be given
consideration.
The next possible single-representation solution to Eq.~\rhoform\
is $T=3$, $Y=4$.  Normally it is discarded because of its complexity.
The simplest representation beyond the doublet is
a Higgs triplet, $T=1$.  In order to have a neutral member the only
possible hypercharge values are $Y=0,\pm 2$.
As we see from Eq.~\rhoform,
$\rho=1$ is not automatic in such a case. Various possibilities for
obtaining $\rho\simeq1$ at tree level can be entertained.  First,
it could be that a single triplet occurs in combination with a doublet.
However, since current experimental limitations
on $\delrho$ are in the vicinity of 0.002 (the exact number depends
upon the confidence level criterion), the vacuum expectation values
of the neutral triplet field must be quite small, implying that the triplets
would play little role in EWSB. For example,
including one or the other of the two triplet representations cited above,
one finds that $|V_{1,0}|/|V_{1/2,1}|\sim 0.03$,
$|V_{1,\pm 2}|/|V_{1/2,1}|\sim 0.03$ yields $\rho\simeq 1.002$,
$\rho\simeq .998$, respectively. (Note that in the first case $\delrho>0$,
while in the second $\delrho<0$.)  Of course, it will be important
to determine if a triplet representation is present, regardless of
the vacuum expectation value of its neutral member. But, as we shall see,
for small vacuum expectation value, an $\emem$ collider would not be useful
for probing the triplet.

Thus, let us consider scenarios in which the triplet vacuum expectation
value could be large.  The first such possibility arises if
$\mt$ is very large. In this case the $t-b$ doublet yields a large
positive contribution to $\delrho$ which could by cancelled by the
negative $\delrho$ that would arise from a $T=1$, $Y=\pm2$ complex
triplet representation. Obviously, this would require fine tuning the
triplet vacuum expectation value to achieve $\rho\simeq 1$. In such a scenario,
quite large values of $\mt$
are required for $V_{1,\pm 2}$ to be big enough to allow the triplet to be an
important player in electroweak symmetry breaking. For example, to
compensate for a ratio of $|V_{1,\pm 2}|/|V_{1/2,1}|\sim 0.2$ we would
need $\mt\sim 475\gev$.

A third, and in my opinion the most attractive, possibility is to combine one
doublet Higgs field with one real $(T=1,Y=0)$ and one complex $(T=1,Y=2)$
field. (For a review and references, see Ref.~\hhg.)
If the Higgs potential is adjusted so that it has a custodial $SU(2)$
symmetry at tree-level, then the neutral members of the two triplet
representations have the same vacuum expectation value and, as can
be explicitly verified using Eq.~\rhoform, $\rho=1$ at tree-level.
Denoting $V_{1/2,1}=a/\sqrt 2$ and $V_{1,0}=V_{1,2}=b$, we find
$\mw^2=\quarter g^2v^2$ with $v^2\equiv a^2+8b^2$. The importance
of the triplet fields in electroweak symmetry breaking can be characterized
by $\tan\theta_H\equiv 2\sqrt 2 b/a$. There is nothing to prevent having
$b>>a$,
in which case EWSB would be dominated by the triplet fields.
We will shortly return to the phenomenology of this model
and the important role an $\emem$ collider would play.

However, I should first explain why this one-doublet, two-triplet
model is not as attractive as a pure doublet model, despite automatically
yielding $\rho=1$ at tree-level. The difficulty is an important
new fine tuning problem that arises at one-loop.  It is easily
verified that the hypercharge gauge interactions ($igB(Y/2)$) violate
the custodial $SU(2)$ symmetry.  This implies that the special form of
the Higgs potential required for $V_{1,0}=V_{1,2}=b$ is {\it infinitely}
renormalized.  Thus, the coefficients in the Higgs potential must be
fine-tuned to preserve $\rho=1$ at 1-loop. Details regarding this fine-tuning
\REF\gvwii{J.F. Gunion, R. Vega, and J. Wudka, \prdj{43} (1991) 2322.}
appear in Ref.~\gvwii. In other words, the most general triplet potential
contains $SU(2)_{custodial}$-violating terms, which, even if set equal
to zero at tree-level, will be generated at one-loop.
This is in contrast to the potential for a model with only Higgs doublets
(plus possible singlets), for which it turns out that even the most
general Higgs potential does not contain terms that can violate the
custodial symmetry. For renormalizable theories (which these models are),
this implies that all radiative corrections to $\rho$ must be finite
if only doublets and singlets are present, a very attractive
conclusion given the experimental constraints on $\rho$.

Thus, triplet models are generally not in
favor with theorists.  However, this does not mean that such models
could not be nature's choice. A complete experimental program should
provide for the ability to search for the many new signatures that would
arise if triplet Higgs representations are present.
\REF\gvwi{J.F. Gunion, R. Vega, and J. Wudka, \prdj{42} (1990) 1673.
See also, P. Bamert and Z. Kunszt, \plbj{306} (1993) 335.}
The phenomenology of the one-doublet, two-triplet model described above
is particularly rich.\refmark\gvwi\

At an $\epem$ collider, the most spectacular and characteristic signal
for a triplet model would be the detection of the doubly charged Higgs boson(s)
contained in complex Higgs triplet representation(s). But, the only available
production reaction would be $\epem\rta H^{++}H^{--}$. Even if
adequate machine energy ($\sqrt s\gsim 2m_{H^{++}}$) is available
and the $H^{++}$ can be seen by this means, the $ZH^{++}H^{--}$
and $\gam  H^{++}H^{--}$ couplings involved in the production mechanism
are fixed purely by weak-isospin and charge, and give no hint of
whether the triplet Higgs field(s) play any role in EWSB.
If the $H^{++}$ decays to a (real or virtual) $\wp\wp$ final state,
the presence of this mode would indicate a non-zero value for the vacuum
expectation value of the neutral member of the associated triplet.
But, the likely absence or small branching ratio
of other channels would mean that the magnitude of this
vacuum expectation value would be essentially impossible to extract.
In order to probe the possible role of a triplet field in
EWSB, other processes must be considered.

To set the stage, let us consider the coupling constant
sum rules that must be satisfied in order that the theory be unitary
at high energy.
\Ref\ghw{J.F. Gunion, H.E. Haber, and J. Wudka, \prdj{43} (1991) 904.}
The two most basic ones are:
$$g^2(4\mw^2-3\mz^2c_W^2){\buildrel \rho\simeq 1 \over \simeq}
g^2\mw^2=\sum_k g^2_{\wp\wm H_k^0}-\sum_l g^2_{\wp\wp H_l^{--}}
\,,\eqn\sumrulei$$
and
$${g^2\mz^4c_W^2\over \mw^2}{\buildrel\rho\simeq 1 \over \simeq}g^2\mz^2
=\sum_k g^{\phantom{2}}_{\wp\wm H_k^0}g^{\phantom{2}}_{ZZ H_k^0}
-\sum_l g^2_{\wp Z H_l^-}\,.
\eqn\sumruleii$$
(Here, $c_W\equiv \cos\theta_W$.)
{}From these two equations we see that
even if there exists a MSM-like $H_1^0$ such that $g_{\wp\wm H_1^0}=g\mw$
and $g_{ZZH_1^0}=g\mz/c_W$, there is still room for more $H_k^0$'s with
(1) big $g_{\wp\wm H_k^0}$,  (2) big $g_{ZZH_k^0}$ {\it provided}:
(1) an $H_l^{--}$ exists with $g_{\wp\wp H_l^{--}}\neq 0$;
(2) $g_{\wp Z H_k^-}\neq 0$ for some singly charged Higgs.
Determination that one of these couplings is non-zero would show absolutely
that triplet Higgs representations exist {\it and} that they play an
important role in EWSB.
For instance, in the one-doublet, one-real-triplet, one-complex-triplet
model described above there is a five-plet (under $SU(2)_{custodial}$)
of Higgs bosons with degenerate masses
($H_5^{--}$, $H_5^{-}$, $H_5^0$, $H_5^{+}$, $H_5^{++}$)
and $WW$-couplings specified below:
$$\eqalign{g^{\phantom{2}}_{H_5^+ \wm Z}&=-g\mw s_H/c_W, \quad
  g^{\phantom{2}}_{H_5^{++} \wm\wm}=\sqrt 2 g\mw s_H,\cr
  \phantom{g_{H_5^+ \wm Z}}&
  g^{\phantom{2}}_{H_5^0\wp\wm}=
    -\half g^{\phantom{2}}_{H_5^0 ZZ} c_W^2=g\mw s_H/\sqrt 3\,, \cr}
\eqn\couplings$$
where we have defined $s_H\equiv\sin\theta_H$. If the triplets are
important in EWSB, then $s_H$ is substantial and the couplings in
question are of the same order as that of the MSM Higgs to $\wp\wm,ZZ$.

How can we look for such couplings?
The most direct technique is to look for a production process that
can occur only if a given coupling is present.
At an $\epem$ collider $g_{H_5^+\wm Z}\neq 0$
leads to $\epem\rta e^+ \nu H_5^- $ via $\wm Z$ fusion.
Despite the kinematically favorable fact that only a single $H_5$
must be produced, the $Z$ couples
weakly to the electron, and so this process does not have a very high
rate in practice. A much more dramatic demonstration of the
triplet Higgs bosons' role in EWSB over a larger $m_{H_5}$
mass region would be possible at an $\emem$
collider via observation of $\emem\rta \nu \nu H_5^{--}$, occurring
by $\wm\wm$ fusion.  This has a rate that is fully competitive
with that normally associated with $\wm\wp$ fusion to a MSM Higgs boson.
\refmark\gvwi\
For instance, in units of the standard $R$, one finds a production
rate of $R\gsim 0.1$ for $\sqrt s\gsim 300+m_{H_5}$
if $\sqrt 2 s_H\sim 1$.  This means that a $\sqrt s=500\gev$ $\emem$
collider (with luminosity comparable to that normally assumed for an $\epem$
collider of this energy)
could easily observe this reaction for $m_{H_5^0}$ up to
about $200-250\gev$.  While the mass reach is only slightly better
than is achievable
via $\epem\rta H_5^{--} H_5^{++}$, the $\wm\wm$ fusion reaction probes
the $\wm\wm H_5^{++}$ coupling that is crucial to the role of the five-plet
in electroweak symmetry breaking.

As an aside, let us imagine the following `frustrating' situation
that might arise if only $\epem$ collisions are available.  In the
one-doublet, two-triplet model being considered, there are altogether
three neutral Higgs bosons with coupling to $\wp\wm$. As well as the $H_5^0$,
we have the $H_1^0$ and $H_1^{0\,\prime}$.  If $\tan\theta_H$ is large
(\ie\ the triplets dominate EWSB), then the $H_1^0$ (which for small
$\tan\theta_H$ plays the role of the MSM Higgs boson) has small
coupling to $\wp\wm$ and in $\epem$ collisions one would see
only the $H_5^0$ and $H_1^{0\,\prime}$ in $\wp\wm$ fusion processes.
The former has $g_{H_5^0\wp\wm}^2=\third g_{H_{MSM}\wp\wm}^2$,
and the latter $g_{H_1^{0\,\prime}\wp\wm}^2=\eightthirds g_{H_{MSM}\wp\wm}^2$,
where $g_{H_{MSM}\wp\wm}^{\phantom{2}}=g\mw$ denotes the coupling strength of
the MSM Higgs boson to $\wp\wm$.
Quite possibly (depending on the masses of the $H_5^0$ and
$H_1^{0\,\prime}$) only the latter would yield a visible rate.
One would observe a neutral Higgs boson, but with a
production cross section much larger than expected
in the Standard Model. Without the $\emem$ collision option, it would be
impossible to do more than guess at the full Higgs sector structure.

Finally, let me briefly review the
decays of the doubly-charged Higgs bosons of the model.
If $s_H$ is not small, the two-body decay $H_5^{--}\rta \wm\wm$
would almost certainly be dominant if kinematically allowed.
Other possible two-body final states
are $H_3^-{\wm}$ and $H_3^-H_3^-$ (where $H_3^-$ is a member of a
surviving $SU(2)_{custodial}$-triplet Higgs species in the model).
\foot{The $H_3^-$ decays primarily to the
heaviest allowed fermion--anti-fermion pair channel.}
If $s_H$ is very small, the $H_3$ modes might dominate if
kinematically allowed. If all these two-body
decays are forbidden, then three-body decays $H_5^{--}\rta \wm
{\wm}^*$ (where ${\wm}^*\rta \ell^-\nu$, \eg) and/or
$H_3^-{\wm}^*$ would be dominant.
\foot{The virtual $H_3^-$ option is small due to the $H_3^-\rta f\anti f$
coupling being proportional to $m_f/\mw$ and, hence, small relative
to the $\wm\rta f\anti f$ coupling.}
If no three-body decays are kinematically accessible,
then the ${\wm}^*{\wm}^*$ four-body decay would be the mode of choice
unless $s_H$ were extremely small.
Of course, we should not forget the possibility of a $H_5^{--}\rta
\ell^-\ell^-$ coupling.  If present at a reasonable level, the
consequent decay would be dominant unless the two-body channels
were kinematically allowed.
\foot{Note, however, that it is unlikely that this coupling could
be large enough to yield direct $H_5^{--}$ production in $\emem$
collisions at an observable level. Nonetheless, an observable rate
cannot be ruled out altogether, and such production should be looked
for. It would provide a very dramatic and unique use for $\emem$
collisions.} This summary makes more explicit the point noted earlier.
Because of the exotic charge of the $H_5^{--}$, it can decay to only
a very few channels, each of which has a highly model-dependent strength.
Thus, it would be difficult to quantitatively determine the role of the
triplet Higgs fields in EWSB simply by examining the $H_5^{--}$ decays.

The only significant background to detection of the $H_5^{--}$ in
the $\wm\wm$ \etc\ final states would derive from the irreducibly present
$\wm\wm\rta\wm\wm$ electroweak subprocesses
(\ie\ those present whether or not there is a Higgs boson).
For a light $H_5^{--}$ resonance, this background would present no problem.
If the $H_5^{--}$ is heavy, a more general discussion of
$\wm\wm$ scattering is appropriate, and is my next topic.

\smallskip
\noindent{\bf 3. Strong $\bf \wm_L\wm_L$ Scattering}
\smallskip

There is no guarantee that the EWSB sector will be entirely
perturbative, or even partially perturbative.  For instance, if there are
no Higgs bosons then we must consider the possibility that
$WW$ scattering becomes strong in all channels and that a
perturbative approach is not possible.  Perhaps a technicolor model
will prove to be correct, perhaps some other approach.
In any case, at best a very incomplete picture of $WW$ interactions will
be possible if only $\epem$ collisions and not $\emem$ collisions are
available.  In general, a full understanding of EWSB will emerge
only if the means by which unitarity is restored at high energy
can be explored in all $WW$ channels, including the $\wm\wm$ channel.

The signal for a strongly interacting $WW$ sector is
non-unitary growth of the $W_LW_L\rta W_LW_L$
scattering amplitudes for longitudinally polarized
gauge bosons at high energy.  Higgs bosons
are required to cancel the bad high energy growth that occurs if only
gauge-boson-exchange electroweak graphs are present.  The single Higgs
boson of the MSM is sufficient to achieve this cancellation in all
$WW$ scattering channels.  If the Higgs boson is not present or is
very heavy, then every $W_LW_L$ scattering channel will become
strongly interacting.  But this is only the simplest of many possibilities.
To illustrate the possible complexity of unitarity cancellations,
and the importance of the $\wm\wm$ channel in being able to fully
explore a scenario in which $WW$ scattering becomes strong,
we return again to the one-doublet,
two-triplet model considered earlier.  Aside from the purely electroweak
graphs, numerous Higgs exchange graphs contribute
to $\wm_L\wm_L\rta \wm_L\wm_L$. These can be divided into:
\pointbegin exchanges of $H_1^0$, $H_1^{0\,\prime}$ and $H_5^0$
in the $t$-channel and $u$-channel; and
\point exchange of $H_5^{--}$ in the $s$-channel.

\noindent
Relative to MSM strength $g^2\mw^2$, the `effective'
(\ie\ after taking into account the signs of the high
energy amplitudes) strengths of these respective graphs
are $c_H^2$, $\eightthirds s_H^2$, $\third s_H^2$, and $-2s_H^2$.
As required, these strengths sum to the MSM result:
$$\sum=g^2\mw^2\left[c_H^2+\left(\eightthirds+\third-2
\right)s_H^2\right]=g^2\mw^2.$$
But, of course, some of the Higgs might be light, and some substantially
heavier, and the manner in which unitarity cancellations occur could be
quite complex.  I will illustrate using two extreme possibilities.

First, suppose $\tan\theta_H\sim 1$, so that $c_H\sim s_H$, and
that $H_1^0$ is light whereas the other Higgs bosons are heavy.
$t$- and $u$-channel exchanges of the $H_1^0$,
with positive weight (the same sign of weight as for
the single Higgs of the MSM) of $c_H^2g^2\mw^2\sim \half g^2\mw^2$,
would partially accomplish the
required unitarity cancellation at a low energy, leaving behind
a unitarity violating high energy behavior that would be weaker
than that found in the MSM when its single Higgs boson is heavy.
The bad high energy behavior in such a case would not
need to be cured until $\sqrt s$ is substantially above the canonical
$\sqrt s\sim 1.8 \tev$ limit found in the MSM single-Higgs
scenario.

In another extreme, suppose the (degenerate) $H_5$ Higgs bosons are light,
and the others substantially heavier.  In net, the $H_5$ contributions
to $\wm_L\wm_L$ scattering have a weight of $-\fivethirds g^2\mw^2$, \ie\
large and opposite in sign to the MSM Higgs.  This would imply
a very rapid high energy growth of the $\wm_L\wm_L$
scattering amplitude until the other Higgs exchanges entered.
In a `hybrid' model, it could happen that there are light $H_5$ Higgs bosons
as described by the model being considered, but that the other
Higgs bosons are not even present.
Then, new physics would have to enter at very low energies in order to
cure the unitarity-violating high energy growth in the $\wm_L\wm_L$ scattering
channel deriving from electroweak and $H_5$-exchange graphs.

Of course, correlated phenomena would be taking place in the
$\wp\wm$ and $\wpm Z$ channels. But, because of the multiplicity
of the different Higgs bosons that might or might not be heavy
(especially if in some more arbitrary scenario
the degeneracy of the $H_5$'s is broken)
the $\wm\wm$ channel would absolutely
be needed in order to fully understand the physics of the EWSB sector.

Certainly, high $\sqrt s$ for the NLC will be
essential to explore EWSB if $WW$ scattering unitary
behavior is only fully achieved at $WW$ energies in the TeV range.
This is the arena for what would probably be a second generation linear
collider with $\sqrt s$ in the $2-4\tev$ range.
If a light MSM-like Higgs boson is not found at
a $\sqrt s\lsim 1 \tev$ $\epem$ collider, then a strong $WW$
scattering scenario becomes likely.  The above illustrations show that
a full understanding of such a sector
will almost certainly require the ability to study $\wm\wm$ scattering
as well as $\wp\wm$, $\wp Z$ and $\wm Z$ scattering. While the latter
three processes are accessible in $\epem$ collisions, only $\emem$
collisions allow a study of the first.  Thus, if one arrives at a juncture
where a multi-TeV linear collider is being considered, and the nature
of EWSB has not been fully resolved at a lower energy machine,
it is especially crucial that an $\emem$ collision option be included
in the machine design.

\smallskip
\noindent{\bf 4. Conclusions}
\smallskip

If the MSM or the minimal supersymmetric model (MSSM) is correct, then
only doublet or lower Higgs representations occur and, in addition,
light Higgs boson(s) should be regarded as likely. Any such light
Higgs is almost certain to be discoverable at an $\epem$ machine with
$\sqrt s\lsim 500\gev-1\tev$. However, it could happen
that anomalies in the $WW$ scattering sector will be observed at
such a machine, either because more complicated Higgs
representations occur and/or because the $WW$ scattering sector becomes
strongly interacting at high energy. In this case, $\emem$ collisions are
almost certain to be required in order to obtain a comprehensive
picture of the nature of electroweak symmetry breaking.
Thus, if at all possible, an option for $\emem$ collisions at the
next linear collider, and especially its second-generation
successor or extension, should be retained.

\smallskip\noindent{\bf 5. Acknowledgements} \smallskip
This review has been supported in part by Department of Energy
grant \#DE-FG03-91ER40674
and by Texas National Research Laboratory grant \#RGFY93-330.
I would like to thank the Aspen Center for Physics for support
during its preparation.  I also gratefully acknowledge
the contributions of my many collaborators to the content of this
report.

\smallskip
\refout
\end